\documentclass[12pt]{article}

\usepackage{graphicx}
\usepackage{times}
\usepackage{bm}
\usepackage{amsmath}
\usepackage{amssymb}
\usepackage{color}
\usepackage{ulem}

\newcommand{\rmi}{{\rm i}}

\newenvironment{sciabstract}{%
\begin{quote} \bf}
{\end{quote}}

\title{Routing the emission of a near-surface light source by a magnetic field}

\author
{F.~Spitzer,$^{1}$ A.N.~Poddubny,$^{2,3,\ast}$ I.A.~Akimov$^{1,2,\ast}$, V.F.~Sapega$^{2}$, \\
L.~Klompmaker$^{1}$, L.E.~Kreilkamp$^{1}$, L.V.~Litvin$^{4}$, R.~Jede$^{4}$, \\
G.~Karczewski$^{5}$, M. Wiater$^{5}$, T.~Wojtowicz$^{5,6}$, D.R.~Yakovlev,$^{1,2}$ and M.~Bayer$^{1,2}$
\\
\\
\normalsize{$^{1}$Experimentelle Physik 2, Technische Universit\"at Dortmund, 44221 Dortmund, Germany}\\
\normalsize{$^{2}$Ioffe Institute, Russian Academy of Sciences, 194021 St. Petersburg, Russia}\\
\normalsize{$^{3}$ITMO University, 197101 St. Petersburg, Russia}\\
\normalsize{$^{4}$Raith GmbH, Konrad-Adenauer-Allee 8, 44263 Dortmund, Germany}\\
\normalsize{$^{5}$Institute of Physics, Polish Academy of Sciences, PL-02668 Warsaw, Poland}\\
\normalsize{$^{6}$International Research Centre MagTop, PL-02668 Warsaw, Poland}\\
\\
\normalsize{$^\ast$To whom correspondence should be addressed;}\\
\normalsize{E-mail: poddubny@coherent.ioffe.ru, ilja.akimov@tu-dortmund.de }
}

\date{}

\begin{document}


\baselineskip24pt


\maketitle


\begin{sciabstract}
Magneto-optical phenomena such as the Faraday and Kerr effects play a decisive role for establishing control over polarization and intensity of optical fields propagating through a medium. Intensity effects where the direction of light emission depends on the orientation of the external magnetic field are of particular interest as they can be used for routing the light. We report on a new class of transverse emission phenomena for light sources located in the vicinity of a surface, where directionality is established perpendicularly to the externally applied magnetic field. We demonstrate the routing of emission for excitons in a diluted-magnetic-semiconductor quantum well. The directionality is significantly enhanced in hybrid plasmonic semiconductor structures due to the generation of plasmonic spin fluxes at the metal-semiconductor interface.
\end{sciabstract}

Control of light emission intensity and propagation direction is required in many domains of modern optics, from macroscopic searchlights to nanoscopic antennas~\cite{Benson-2011, Lodahl-et-al-2017}.
The use of magnetic forces for routing emission is highly appealing for applications in nanophotonic circuits, magneto-optical storage, and precision metrology~\cite{Temnov-2010, Akimov-2012, Armelles-2012, Bossini-2016}.
In particular magneto-optical phenomena such as the Faraday (Kerr) effect are widely used to control the polarization of transmitted (reflected) light rays~\cite{Zvezdin-TMOKE,MO-book}.
Significant progress in the enhancement of these magneto-optical effects has been achieved recently by combination of magnetic and plasmonic materials~\cite{Belotelov-2009, Chin-2013, Kreilkamp-2013, Floess-2017}.
However, the magnetic field control over the emitter's directionality requires dedicated tailoring of the structures for implementation in order to meet the required conditions. Until now it has been demonstrated only for emission of chiral objects in Faraday geometry, along the axis parallel to the magnetic field~\cite{MChD97}. Transverse intensity effects in emission with directionality perpendicular to the magnetic field, have not been reported yet.

\begin{figure}[t]
\centering
\includegraphics[width=.92\textwidth]{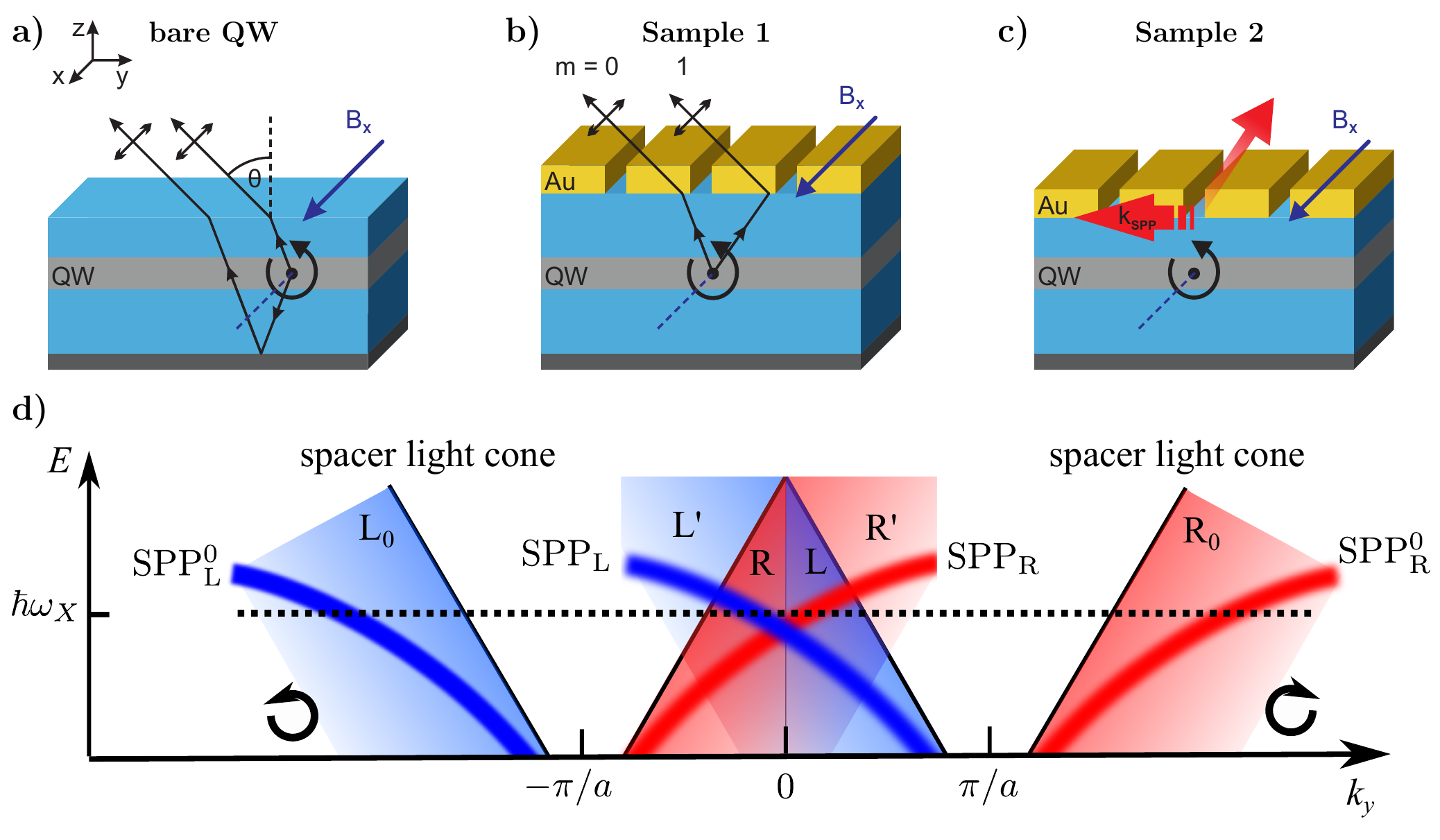}
\caption{\textbf{Exciton-mediated optical spin fluxes routed by magnetic field.}\newline
Top panels (a)-(c) show the main mechanisms which lead to transverse magnetic routing of light emission (TMRLE): (a) Interference of direct and backside-reflected beams (solid lines), occuring also for a bare QW structure; (b) Diffraction at the grating causes interference of the beams corresponding to different orders ($m = 0,1$), which takes place even for thick spacers; (c) Emission into surface plasmon-polariton (SPP) modes if the QW is located close enough to the metal-semiconductor interface. (d) Schematic presentation of dispersion for evanescent electromagnetic waves in the plasmonic grating of (b,c) and their contribution to TMRLE in the far field emission. The dispersion branches of the SPPs at the homogeneous gold boundary are $\rm SPP^{0}_{R,L}$ and the diffracted dispersion branches are SPP$_{\rm R}$ and SPP$_{\rm L}$.}
\label{Fig:scheme}
\end{figure}

Here, we demonstrate transverse magnetic routing of light emission (TMRLE). It is observed when the light source coupling to a magnetic field $\bm B$ is located in the vicinity of a surface that breaks the mirror symmetry of the medium. Namely, the wave vector of the emission that points in the routing direction $\bm k$ is proportional to
\begin{equation}
\bm k\propto \bm B\times \bm e_{z} \label{eq:kSPP}
\end{equation}
where $z$ is the direction normal to the surface, $\bm e_{z}$ is the unit vector along $z$, and $\bm B\parallel x$ as shown in Fig.~\ref{Fig:scheme}.

TMRLE requires two key features to be fulfilled: First, the optical selection rules of the light source need to be modified by the magnetic field, which is an intrinsic property of any emitter in magnetic materials. Second, the emitted light should have non-zero transverse spin (angular momentum) $\bm S\parallel x \perp \bm k$.
Non-zero local transverse spin is present in any structure with broken $z\to -z$ reflection symmetry, for instance, when the emitter is placed near a planar mirror~\cite{Bliokh-2015}. Conventional electromagnetic plane waves remain linearly polarized on average and their propagation direction is independent of their polarization. Therefore, only weak directionality effects can be expected in bulk material.
On the contrary, subwavelength optical fields possess strong transverse spin locked to their propagation direction, i.e. spin fluxes~\cite{Lodahl-et-al-2017, Glazov-2007, Bliokh-et-al-2015}. This enables fascinating phenomena such as lateral pressure controlled by the polarization of light \cite{Zayatz-2015}, photonic spin Hall effect in hyperbolic materials~\cite{Poddubny-2014} and photonic wheels in tightly focused beams~\cite{Leuchs-rev-2015}. Generation of photonic spin fluxes has been established in dielectric photonic structures containing a single atom~\cite{Rauschenbeutel-2013,Dayan-2014} or a semiconductor quantum dot~\cite{Oulton-2013, Lodahl-2015, Makhonin-2016}.
For plasmonic excitations, which are based on collective oscillations of electrons in a metal, spin fluxes have also been visualized via plasmon scattering on nanoparticles~\cite{Capasso-2013, Zayats-2013, Leuchs-2013, Zayats-2014, Rauschenbeutel-2014, Martinez-2014, Grosjean-2015, Iorsh-2017}. Therefore, the use of subwavelength optical fields, which are present in photonic as well as plasmonic nanostructures, is expected to increase TMRLE.

For demonstration of TMRLE we consider the exciton emission from a semiconductor quantum well (QW).
In order to assess the importance of subwavelength optical fields we also study hybrid plasmonic structures where the metallic grating is placed at the top of the semiconductor structure in close proximity to the QW. The metallic grating contributes in two ways: (i) The metal-semiconductor interface supports surface plasmon polaritons (SPPs); (ii) The periodic grating allows detection of optical spin fluxes carried by surface modes in the far field radiation. Here, we concentrate on three prominent cases, which are shown in Fig.~\ref{Fig:scheme}: a {\it bare QW} and two structures covered with identical one-dimensional plasmonic gratings with a period $a$ but different spacer thicknesses between the grating and the QW.

The QW emission originates from the lowest energy states which correspond to optically active heavy hole excitons. Due to QW inhomogeneity the excitons are localized and can be modeled as point electric dipoles. In zero magnetic field $\bm{B}$ the excitons have angular momentum projection $J_z = \pm 1$ along the QW confinement $z$ axis, corresponding to circularly polarized dipoles that rotate in the $xy$-plane clockwise or counter-clockwise, $d_{x}\bm{e}_x\mp \rmi d_{y}\bm{e}_y$ with $d_x=d_y$. Therefore, at $B=0$ light is emitted with equal probability in the lateral directions.

An in-plane magnetic field $\bm{B} \parallel x$ modifies the selection rules of the optical transitions, that become elliptically polarized in the $yz$-plane corresponding to dipoles with $\bm {d}_X^\pm = d_{y}\bm e_{y} \mp \rmi d_z\bm e_{z}$. The circular polarization degree is $P_c = \pm 2d_{y}d_{z} / (d_{y}^2 + d_{z}^2) \approx \pm \frac{2}{3}\Delta_{h,F}/\Delta_{lh}$ for $\Delta_{h,F} \ll \Delta_{lh}$. Here, $\Delta_{h,F}$ is the Zeeman splitting of the heavy holes in the Faraday geometry, which grows linearly with $B$, and $\Delta_{lh}$ is the energy splitting between the heavy and light hole states due to QW confinement and strain (for details see sections 1 and 2 of Supplementing Material~\cite{supplement}). Due to the lack of mirror symmetry along the $z$ axis the optical field carries non-zero transverse spin in all structures shown in Fig.~\ref{Fig:scheme}. And therefore, the circularly polarized excitons preferentially couple either to left- or to right-going waves, depending on the magnetic field sign. As a result, the emission becomes directional which is the main focus of this work.

It is important to distinguish between far-field and near-field effects. The first ones can be considered as an interference of several emitted light rays, while the latter ones are initiated by exciton emission into the evanescent (sub-wavelength) optical modes which exist at the semiconductor-metal interface. Fig.~\ref{Fig:scheme}(a) and (b) demonstrate the far field effects. In (a) interference takes place between the directly emitted and the reflected electromagnetic waves. The reflection occurs at the back side of the structure. In (b) the interference occurs between the zero and first order diffracted beams which are transmitted through the grating. The interfering beams originate from the same dipole and propagate along the same direction in free space outside the sample. For a given emission angle $\theta$, which is defined with respect to the $z$ axis, and a given optical frequency of the exciton resonance $\omega_X$ the phase between the interfering tilted beams is determined by the helicity of the dipole in the $yz$-plane and therefore directionality of the QW emission arises in an applied magnetic field.

The directionality can be enhanced in the near field when the QW layer is located in close proximity of a semiconductor-metal interface and the excitons emit into evanescent waves, see Fig.~\ref{Fig:scheme}(c). At a given frequency these can be evanescent photon modes with arbitrary in-plane wave vector $k_{y}$ or SPP waves, where the value of $k_{y}$ is fixed by the dispersion relation. In both cases the evanescent TM-polarized waves carry transverse spin $S_{x}\propto \rmi [\bm E\times \bm E^{*}]_{x}$, where the electric field is $\bm E\propto k_{z}\bm e_{y}-k_{y}\bm e_{z} $ and $k_{z}$ is imaginary. Hence, for the right-going evanescent waves $S_{x}>0$ and for the left-going waves $S_{x}<0$~\cite{Bliokh-et-al-2015}. This is illustrated by the colored  areas R$_{0}$ and L$_{0}$ in the dispersion diagram presented in Fig.~\ref{Fig:scheme}(d). Due to the modified exciton selection rules, the generated plasmon spin is pinned to the magnetic field, $ \bm S\parallel \bm B \label{eq:S}\:.$ This results in magnetic-field-induced photonic as well as plasmonic spin fluxes.

The grating enables outcoupling of the evanescent waves into the far field. This can be described by translating the regions R$_0$ and L$_0$ in Fig.~\ref{Fig:scheme}(d) by the reciprocal lattice vector $k_{y}=-2\pi/a$ and $k_{y}=2\pi/a$, respectively, which leads to the areas R, R$'$, L, L$'$. Interestingly, in periodic structures the directionality of the transverse photon and plasmon spin fluxes observed in the far field can compensate or enhance each other.  The spin generated for a given frequency as function of $k_{y}$ has maxima (i) at the SPP branches and (ii) in the regions just below the light lines, where $|k_{z}|$ is small and the electric field $E\propto{\rm e}^{-|k_{z}|z}$ is rather large.  The regions corresponding to diffracted right- and left-propagating evanescent waves overlap with each other.  As a result, the transverse spin sign reverses at the boundaries L$'-$R and L$-$R$'$ of the central light cone. In the regions L$'$ (R$'$) the routing effect becomes larger because the evanescent wave contribution from the region R (L), having opposite sign, vanishes. The relative sign of the diffracted plasmon and photon transverse spin inside the light cone is energy-dependent. Below the crossing point of the SPP$_{\rm R}$ and SPP$_{\rm L}$ branches the signs are the same, above the crossing point they are opposite.

\begin{figure}[t!]
\centering
\includegraphics[width=.91\textwidth]{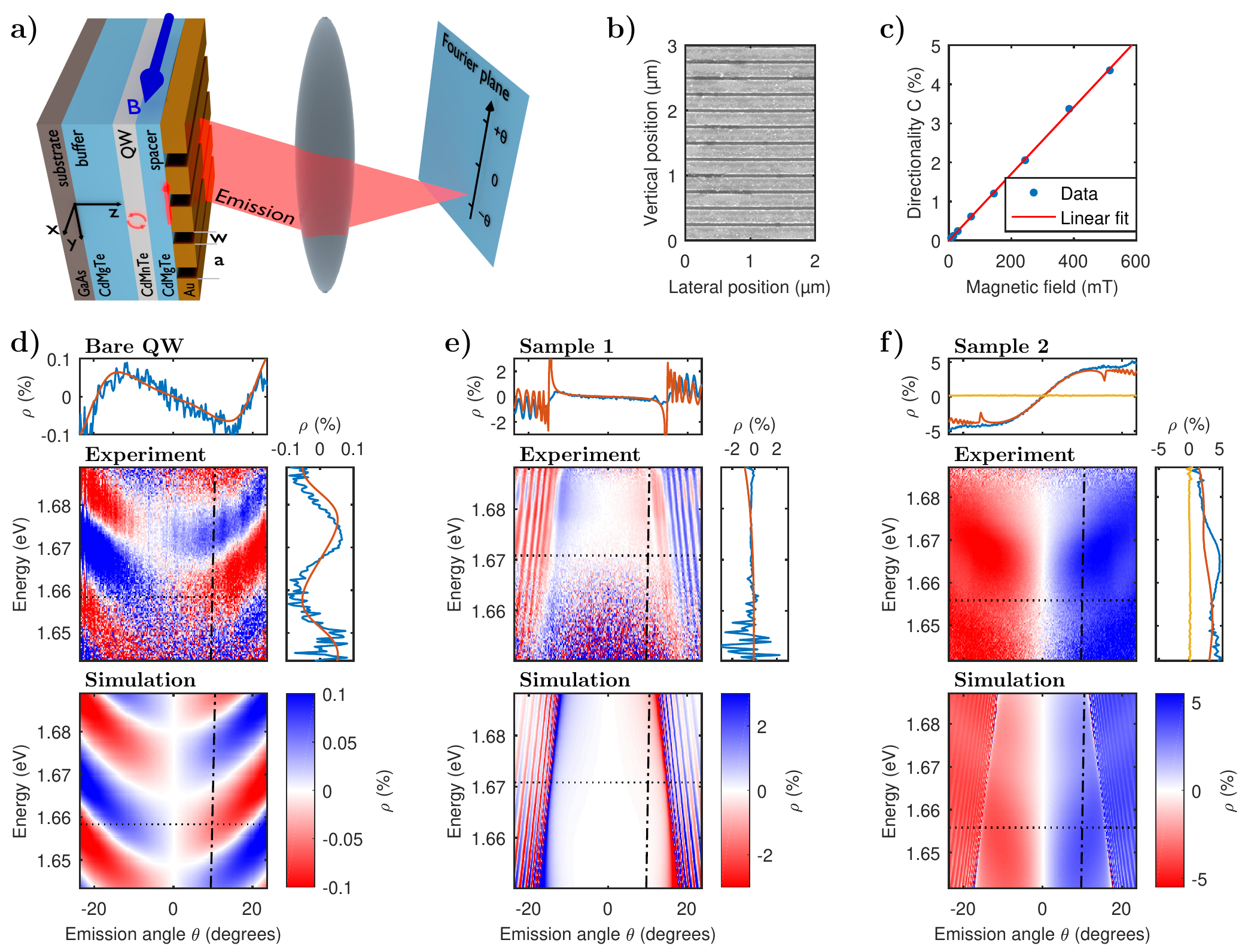}
\caption{\textbf{Demonstration of transverse magnetic routing of light emission (TMRLE).}\newline
A Fourier imaging setup is used to convert the angular dependence of emitted light into spatial dependence in the Fourier plane which is projected onto the spectrometer slit. Angle- and spectrally-resolved exciton emission is detected as a contour pattern with a two-dimensional charge-coupled device (CCD) matrix detector, attached to an imaging spectrometer. (b) Scanning electron microscope (SEM) image of the plasmonic grating with $250\,$nm period (Sample 2). (c) Magnetic field dependence of the TMRLE directionality factor $C$ measured in sample 2 at a photon energy $1.669\,$eV for $\theta=10^\circ$. The corresponding patterns of the magnetic field induced variation of the photoluminescence (PL) intensity $\rho(\hbar\omega,\theta)$ are presented in (d) for the bare QW, in (e) for sample 1 and in (f) for sample 2. Upper and lower colored panels correspond to measured and calculated $\rho(\hbar\omega, \theta)$-patterns, respectively. Note the different color scales in these panels. Side plots at the measured patterns show cross-sections along fixed photon energy $\hbar\omega$ as indicated by the dotted line (upper plot) or fixed angle $\theta$ as indicated by the dash-dotted line (right plot). Blue and red curves in these plots correspond to cross-sections of experimental data and calculation results, respectively. Yellow curves in the side plots of (f) show PL signal measured in $s-$polarization. All measurements are performed at temperature $T=10\,$K and magnetic field $B=520\,$mT.}
\label{Fig:PLMOKE}
\end{figure}

The experimental demonstration of TMRLE is presented in Fig.~\ref{Fig:PLMOKE}. To that end we have chosen diluted-magnetic-semiconductor  Cd$_{0.95}$Mn$_{0.05}$Te/Cd$_{0.73}$Mg$_{0.27}$Te QW structures as light sources, which represent model systems demonstrating bright exciton emission with well established parameters~\cite{Furdyna, Kossut-book}. The strong $p-d$ exchange interaction of valence band holes with magnetic Mn$^{2+}$ ions leads to the giant Zeeman splitting of the hole spin levels~\cite{Furdyna, Kossut-book, DMS-1994}, facilitating a high circular polarization degree $P_c$. For QW width of 10~nm and the parameters used in our experiment (temperature 10~K and $B=$520~mT), $\Delta_{h,F} \approx 4$~meV and $\Delta_{lh} = 20$~meV, so that $P_c\approx0.13$ can be expected (see section 1 of Supplementing Material~\cite{supplement}).

Rectangular gold gratings with size of $200\times200~\mu{\rm m}^2$ were patterned using electron beam lithography and subsequent lift-off (see section 3 of Supplementing Material~\cite{supplement}). The grating period of 250~nm and the slit width of 55~nm were evaluated from the scanning electron microscope (SEM) image shown in Fig.~\ref{Fig:PLMOKE}(b). The thickness of the Au layer is about 45~nm. The studied samples contain also areas without gold stripes at the top, i.e. corresponding to the bare QW. Two samples from structures grown with spacer thicknesses of 250~nm ({\it sample 1}) and 32~nm ({\it sample 2}) between the grating and the QW were fabricated. The parameters of the gratings were chosen in order to obtain far-field emission of the SPPs from the semiconductor-gold interface at optical frequencies $\omega_{\rm SPP}\approx\omega_X$ in a small range of angles $\theta$ as confirmed by reflectivity spectra (see section 4 of Supplementing Material~\cite{supplement}).

For optical studies we use a Fourier imaging setup in combination with a spectrometer, which allows us to acquire both the angular and the spectral dependence of the emitted light intensity in a single acquisition (see Fig.~\ref{Fig:PLMOKE}(a)). We use laser excitation with a photon energy of 2.25~eV in order to populate the QW with excitons. The emission band of QW excitons is centered at energy $\hbar\omega \approx 1.67$~eV and depends weakly on the emission angle $\theta$. The spectral width at half maximum of the PL band of about 10~meV allows us to measure the emission pattern in the spectral range from 1.64 to 1.69~eV.

In the magneto-optical measurements the angle-resolved photoluminescence (PL) is detected in the $yz-$plane perpendicular to the magnetic field $\bm{B} \parallel x$ (see Fig.~\ref{Fig:PLMOKE}(a)). The magnetic-field-induced variation of the PL intensity is assessed by the normalized difference
\begin{equation}
\rho(B) = \frac{I(+B)- I(-B)}{I(+B)+ I(-B)}\:,
\label{eq:rho}
\end{equation}
where $I(+B)$ and $I(-B)$ are the intensity patterns measured for opposite directions of the magnetic field $B$. The antisymmetric part of $\rho$ with respect to the emission angle $\theta$ defines the directionality factor $C(\theta) = [\rho(\theta)-\rho(-\theta)]/2$.

Contour plots of $\rho(\hbar\omega,\theta)$ for $B=520$~mT and $p$-polarized detection are shown for the bare QW with 32~nm thick spacer as well as for sample 1 and 2 in panels (d)-(f) of Fig.~\ref{Fig:PLMOKE}. A variation of the QW emission takes place in all three cases. However, there are significant differences in the angular and spectral distribution as well as in the magnitude of $\rho$. All images show that $\rho$ is an odd function with respect to the emission angle, i.e. $\rho(\hbar\omega,\theta)=-\rho(\hbar\omega,-\theta)$. Thus, in our case the magnetic-field-induced variation of the PL intensity is fully determined by the TMRLE with directionality $C(\hbar\omega,\theta)=\rho(\hbar\omega,\theta)$. In full accord with the theoretical predictions the strength of the effect follows a linear dependence on magnetic field (Fig.~\ref{Fig:PLMOKE}(c)). Note, that the PL spectra from all structures contain also emission from the GaAs substrate centered at photon energies of 1.49~eV and 1.51~eV. These emission bands originate from non-magnetic constituents and do not show any intensity variation within the accuracy of the experiment ($\rho < 10^{-4}$). The directionality of the emission is not observed for $s$-polarized emission, which is in agreement with our TMRLE analysis requiring two orthogonal electric field components in the $yz$-emission plane (see yellow curves in the insets of Fig.~\ref{Fig:PLMOKE}(f)).

Now we discuss the main features of the TMRLE effect for the different structures. In the bare QW sample the magnetic-field-induced changes of the QW emission are very small, $|\rho| <0.1\%$ (see Fig.~\ref{Fig:PLMOKE}(d)). Further, there is a strong frequency dependence, $\rho(\hbar\omega)$ oscillates around zero with a period of about 30~meV. For this reason the overall directionality decreases even further after averaging over the photon energy. As discussed above this behaviour occurs due to interference of the direct and the back-reflected emission beams (see Fig.~\ref{Fig:scheme}(a)) and does not require coupling to subwavelength optical fields. Another far-field contribution is manifested in sample 1 with the large 250~nm thick spacer and the grating at the top. The step-like increase of $\rho$ at the threshold angle $\theta_c\approx 11^\circ-17^\circ$ in Fig.~\ref{Fig:PLMOKE}(e) is correlated with the appearance of the first order diffracted beam for the emitted light as shown in Fig.~\ref{Fig:scheme}(b). Simultaneously, an interference from the waves reflected from the sample backside contributes, similar to that visible in Fig.~\ref{Fig:PLMOKE}(d), which explains the additional interference fringes at $\theta>\theta_c$ in Fig.~\ref{Fig:PLMOKE}(e).

The strongest TMRLE is observed in sample 2 with the thin spacer between the QW and the gold grating, where the exciton emission is partly transferred to SPPs. In Fig.~\ref{Fig:PLMOKE}(f) one observes a strong directionality on the order of $5\%$. The spectral dependence is weak as it is determined by the resonance condition $\omega_X = \omega_{\rm SPP}$ (Fig.~\ref{Fig:scheme}(d)). Since the bandwidth of the SPP resonance ($\sim 50\,$meV) is significantly larger than the spectral width of the exciton PL band ($\sim 10\,$meV) no strong spectral dependence of $\rho$ is observed. Also weak oscillations are superimposed on the main signal due to the interference fringes discussed above. However, they do not diminish the effect but $\rho$ remains positive for $\theta>0$.

The grating structures show an additional interesting behaviour, which is attributed to the relative contribution of plasmonic and photonic spin fluxes discussed in Fig.~\ref{Fig:scheme}(d). When the spacer thickness increases from 32 to 250~nm the SPP contribution to the directionality diminishes. This is due to the exponential decay of the evanescent SPP wave into the QW layer and the corresponding decrease of the interaction between SPPs and excitons. However, the routing effect is still present and its sign is opposite to that observed in sample 2. From Fig.~\ref{Fig:PLMOKE}(e) we find $\rho\sim-0.5\%$ in sample 1 for small emission angles $0<\theta < +10^\circ$. As discussed above the relative sign of the plasmon transverse spin and the transverse spin of the nonresonant wave is energy-dependent. Below the crossing point of the SPP$_{\rm R}$ and SPP$_{\rm L}$ branches the signs are the same, and above the crossing point they are opposite. Our experimental situation corresponds to the case when the exciton frequency is slightly above the crossing point, see the dotted line in Fig.~\ref{Fig:scheme}(d). Hence, when the spacer thickness increases, the plasmonic effect is suppressed and substituted by the nonresonant one, and the directionality sign reverses, cf. signs in Fig.~\ref{Fig:PLMOKE}(f) and the central part of Fig.~\ref{Fig:PLMOKE}(e).

The theoretical calculations presented in the lower panels of Fig.~\ref{Fig:PLMOKE}(d)-(f) show good agreement with the experimental data (for details see section 5 of Supplementing Material~\cite{supplement}). Here, only the degree of circular polarization $P_c$ was used as a fit parameter. A value of $P_c= 2.6\%$ was used consistently in all the simulations. The decrease of polarization as compared to the value  $P_c = 13\%$ predicted for the idealized QW with infinite walls can be explained by the reduced overlap between the light and heavy hole wavefunctions $|\langle \psi_{lh}|\psi_{hh}\rangle|^{2}\ll 1$ in the actual QW.

In conclusion, we have proposed and demonstrated a new class of transverse emission phenomena, where directional emission from a light source arises perpendicularly to the magnetic field orientation. The proposed routing effects require an emitter placed near a planar surface which breaks the mirror symmetry and introduces a nonzero transverse spin of light, that is locked to the propagation direction and can be controlled by the magnetic field. The experimental demonstration has been achieved for exciton emission from a diluted-magnetic-semiconductor quantum well, which is characterized by a giant Zeeman splitting. The routing is linear in magnetic field, in accordance with the theoretical predictions. When the distance between the emitter and the surface is large, the transverse spin is caused by the far-field interference effect and the routing is weak. The strongest directionality is achieved for a quantum well located several tens of nm apart from a metal-semiconductor interface. At such distance the quantum well is coupled to surface plasmon polaritons, that carry large transverse spin and are efficiently controlled by the magnetic field direction. Our results demonstrate that plasmonic structures, similarly to dielectric ones,  can be efficiently used for readout of transverse spin in semiconductor nanostructures. The current studies have been performed in magnetic structures with giant Zeeman splitting which facilitate strong directionality at low temperatures. However proper choice of materials with intrinsically large Zeeman splittings (e.g. InSb, HgMgTe, novel 2D materials) could be interesting for realization of magnetic field routing even at room temperatures.

\noindent{\bf Acknowledgements}

We are grateful to  M.M. Glazov, E.L. Ivchenko, V.L. Korenev, I.V. Iorsh, A.K. Samusev  for useful discussions. We acknowledge the financial support by the Deutsche Forschungsgemeinschaft through the International Collaborative Research Centre 160 (Project C5).  ANP acknowledges the partial financial support from the Russian Foundation for Basic Research Grants No. 15-52-12012-NNIO\_a and 15-52-12011-NNIO\_a and the "Basis" Foundation. MB acknowledges support by RF Government Grant No. 14.Z50.31.0021. The research in Poland was partially supported by the National Science Centre (Poland) through Grants No. DEC-2012/06/A/ST3/00247 and No. DEC-2014/14/M/ST3/00484, as well as by the Foundation for Polish Science through the IRA Programme co-financed by EU within SG OP.

\vskip 1cm
\noindent{\bf Additional information}

The authors declare no competing financial interests.


\renewcommand{\thefigure}{S\arabic{figure}}
\setcounter{figure}{0}

\renewcommand{\theequation}{S\arabic{equation}}
\setcounter{equation}{0}

\makeatletter
\renewcommand\@bibitem[1]{\item\if@filesw \immediate\write\@auxout
    {\string\bibcite{#1}{S\the\value{\@listctr}}}\fi\ignorespaces}
\def\@biblabel#1{[S#1]}
\makeatother

\begin{large}
\vspace*{\fill}
\center{\bf Supplementing material:\\Routing the emission from a near-surface light source by a magnetic field}
\vspace*{\fill}
\end{large}

\noindent{\bf 1. Giant Zeeman splitting of valence band holes }

In Faraday geometry the exciton energy splitting between the Zeeman levels in the diluted-magnetic-semiconductor Cd$_{1-x}$Mn$_x$Te  is described by the well established formula~\cite{Furdyna, Kossut-book}
\begin{equation}
\label{eq:giantZeeman} \Delta_{Z}(B) = x N_0 (\alpha - \beta) \langle S_{z}^{Mn} (B) \rangle,
\end{equation}
where $x$ is the Mn$^{2+}$ concentration, $N_0 \alpha = 0.22$~eV and $N_0 \beta=-0.88$~eV are the exchange constants for the conduction and valence bands, respectively, and $\langle S_{z}^{Mn} (B) \rangle$ is the thermal average of the Mn spin projection along $B$. This average is given by the modified Brillouin function $\rm B_I$ for $I=5/2$:
\begin{equation}
\label{eq:Brillouin} \langle S_{z}^{Mn} (B) \rangle = S_{eff} \mathrm{B_{5/2}} \left[ \frac{5 \mu_B g_{Mn} B }{2 k_B T_{eff}} \right],
\end{equation}
where $\mu_B$ is the Bohr magneton, $k_B$ is the Boltzmann constant, $T$ is the lattice (bath) temperature and $g_{Mn}=2.01$ is the Mn$^{2+}$ $g$ factor. $S_{eff}$ is the effective spin and $T_{eff} = T + T_0$ is the effective temperature. The parameters $S_{eff}$ and $T_0$ enable a phenomenological description of the antiferromagnetic Mn-Mn exchange interaction.

\begin{figure}[t!]
\begin{center}
\includegraphics[width=0.5\textwidth]{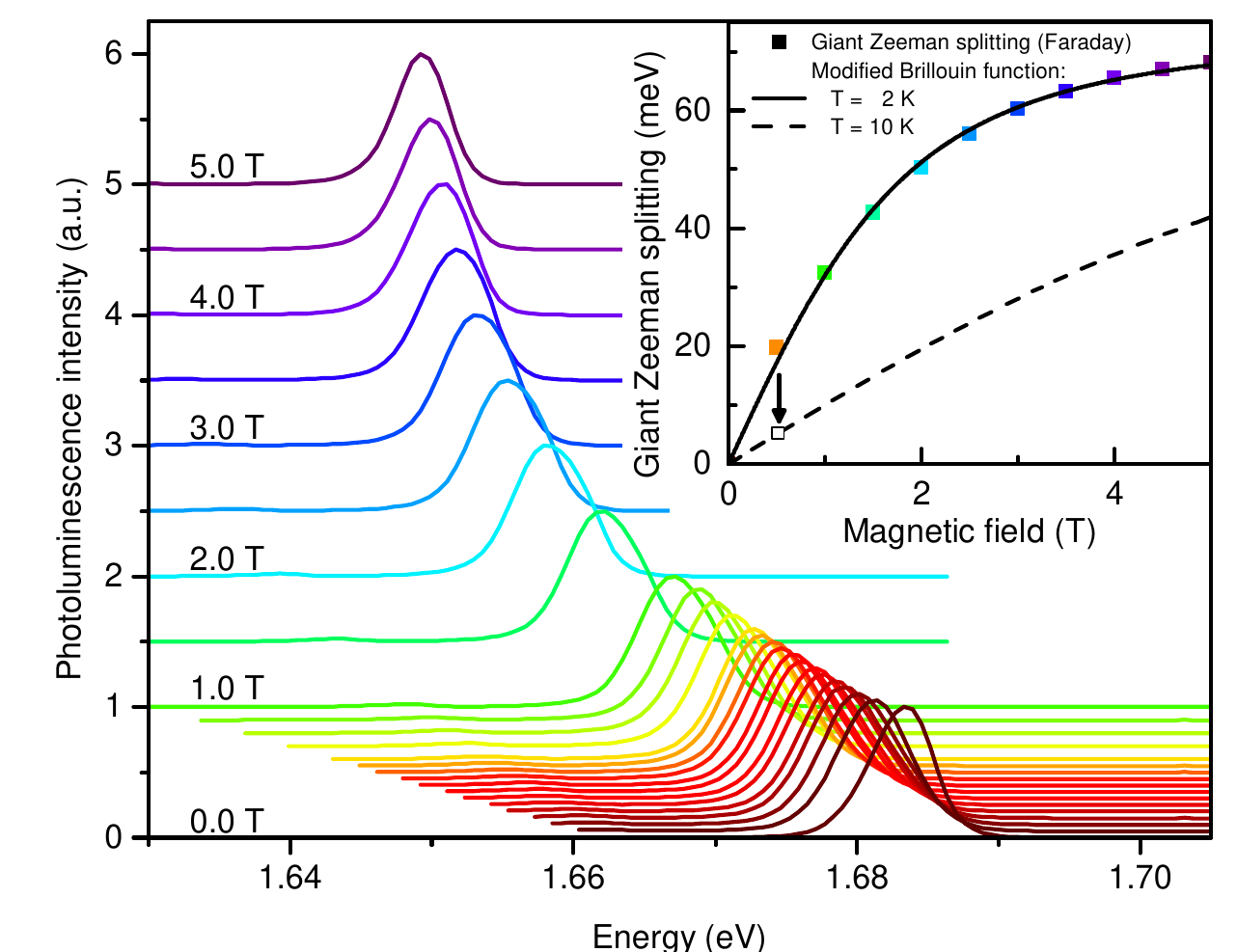}
\end{center}
\caption{Circularly polarized ($\sigma^+$) PL spectra of QW excitons for magnetic fields up to $5\,$T in Faraday geometry at $T = 2\,$K. Inset shows the magnetic field dependence of exciton Zeeman splitting which is determined as twice the shift of the PL peak position with respect to the zero field peak. Solid line is fit with the modified Brillouin function (Eq.~(\ref{eq:giantZeeman})) with $x=0.05$, $S_{eff}=5/2$ and $T_{eff}=3$~K. Dashed line shows the Zeeman splitting obtained from Eq.~(\ref{eq:giantZeeman}) using the same parameters but for increased temperature $T=10\,$K. From this curve, we obtain $\Delta_Z = 5\,$meV at $T=10\,$K and $B=520\,$mT (open  square).}\label{fig:SZeeman}
\end{figure}

Values of the giant Zeeman splitting in the studied structures were obtained from magneto-PL measurements of the bare QW structure. The sample was mounted in the variable temperature insert of a liquid helium bath cryostatat $T=2\,$K. Magnetic fields of up to $5\,$T were applied in Faraday geometry. Excitons were excited by a He-Ne laser with photon energy $1.96$~eV. The emitted light was spectrally dispersed by a double monochromator and detected by a photomultiplier connected to a photon counting unit. The PL spectra measured in $\sigma^{+}$ polarization are shown in Fig.~\ref{fig:SZeeman}.

In order to calculate the giant Zeeman splitting of the heavy hole exciton we take twice the difference of the PL peak position at $B = 0$ and $B > 0$ (filled squares in the inset). From a modified Brillouin function fit according to Eq.~\ref{eq:giantZeeman} (solid line) we obtain the Mn$^{2+}$ concentration of $x = 0.05$ and the effective temperature of $T_{eff} = 3\,$K.

The angle resolved measurements on the plasmonic structures were performed in a flow cryostat at higher temperature of 10~K. The corresponding Zeeman splitting is shown in the inset of Fig.~\ref{fig:SZeeman} for $T = 10\,$K (dashed line). We obtain a total giant Zeeman splitting of $\Delta_Z = 5\,$meV at a magnetic field $B = 520\,$mT (the empty square). The hole contribution to this splitting is $\Delta_{h,F} = \frac{|\beta|}{|\alpha - \beta|} \Delta_Z = 4\,$meV.\\

\newpage

\noindent{\bf 2. Degree of circular polarization of exciton states in magnetic field.}

\begin{figure}[t!]
\begin{center}
\includegraphics[width=0.8\textwidth]{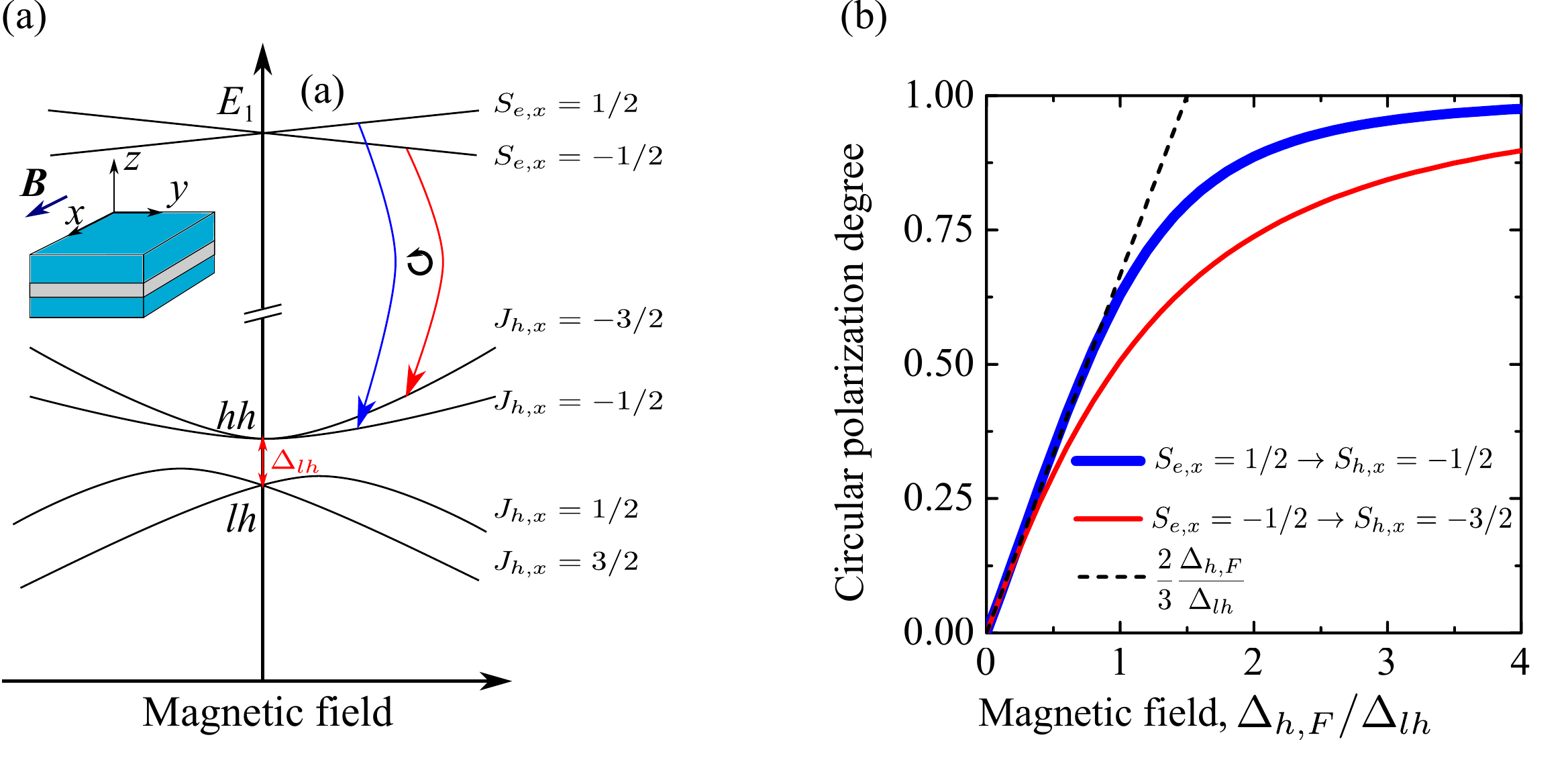}
\end{center}
\caption{(a) Scheme of the Zeeman splitting of the conduction and valence bands in the Voigt geometry.
Red and blue arrows indicate the optically allowed transitions between Zeeman-split ground states in the conduction and valence band; at large magnetic field when $|\Delta_{h,F}|\gg \Delta_{lh}$ they become 100\% circularly polarized.
(b) Calculated dependence of the circular polarization degree $P_{c}$ on the magnetic field strength. Solid blue and red curves correspond to the transitions indicated in (a), the dashed black line shows the approximate result Eq.~\eqref{eq:Pca}.
}\label{fig:SMagn}
\end{figure}

We assume that the confinement of the excitons in the direction $z$ perpendicular to the quantum well plane is stronger than the lateral confinement. Hence, the circular polarization degree can be calculated by generalizing the theory for a magnetic QW developed in Ref.~\cite{DMS-1994}. The scheme of the Zeeman splitting of the conduction and valence band states \ref{fig:SMagn} in the Voigt geometry is shown in Fig.~\ref{fig:SMagn}(a). The $\Gamma_{6}$ conduction band splits similar to the case of Faraday geometry. The eigenstates are $\psi_{e,\uparrow(\downarrow)}=\uparrow(\downarrow)S$ with  energies
 $\pm \tfrac{1}{2}\alpha \langle S_{z}^{Mn} (B) \rangle$, where $s$ is the $s$-type Bloch amplitude and $\uparrow(\downarrow)$ are the spinors with $S_{e,x}=\pm 1/2$. The splitting of the $\Gamma_{8}$ valence band states is described by the following Luttinger Hamiltonian \cite{DMS-1994,Winkler}
\begin{equation}
H_{\uparrow(\downarrow)}=\begin{pmatrix}
-(\gamma_{1}+\gamma_{2})k_{z}^{2}\pm 3\mathcal{B} &
\sqrt{3}\gamma_{2}k_{z}^{2}
\\\sqrt{3}\gamma_{2}k_{z}^{2}&-(\gamma_{1}-\gamma_{2})k_{z}^{2}\mp \mathcal B
\end{pmatrix}\:,\label{eq:HB}
\end{equation}
where the $\uparrow (\downarrow)$ subscripts label the subspaces of the Hamiltonian with $J_{h,x}=[3/2,-1/2]$ and $J_{h,x}=[-3/2,1/2]$ projections of the total angular momentum $\bm J_{h}$ on the magnetic field direction $x$, and $k_{z}\approx \pi/d$ is the effective hole wave vector in $z$ direction ($d$ is the QW width).
The parameter  $\mathcal B=1/6 xN_{0}\beta \langle S_{x}^{Mn} (B) \rangle$ describes the  exchange interaction with the Mn spins. We use the Luttinger parameters $\gamma_{1}\approx 5.3 \hbar^{2}/2m_{0}$ and $\gamma_{2}\approx 1.6 \hbar^{2}/2m_{0}$~\cite{LeSiDang}, where $m_{0}$ is the free electron mass.
 The Hamiltonian Eq.~\eqref{eq:HB} describes the competition between the Zeeman splitting in the magnetic field $\bm B\parallel x$ and the size quantization along the $z$ axis.
At zero magnetic field the hole angular momentum is oriented perpendicular to the QW plane, $\bm J_{h}\parallel z$. The heavy and light holes have quantization energies $E_{hh(lh)}=(\gamma_{1}\pm2\gamma_{2})k_{z}^{2}$. The heavy-light hole splitting in the approximation of infinite barriers, $k_{z}=\pi/d$, and neglecting strain, is given by
\begin{equation}
\Delta_{lh}=|E_{lh}-E_{hh}|=4\gamma_{2}(\pi/d)^{2}\approx 20~\rm meV.\label{eq:DeltaHL}
\end{equation}
In actual QW this value is smaller due to finite height of the barriers. However additional contribution due to strain increases $\Delta_{lh}$~\cite{Strain,Yakovlev-92}. Experimentally we evaluate the value of 20~meV from the reflection spectra, which means that the current estimate for $\Delta_{lh}$ in ~\eqref{eq:DeltaHL} is valid.
The Zeeman splitting of the heavy hole states at weak magnetic fields ($|\mathcal B|\ll \gamma_{1,2}k_{z}^{2}$) is given by
\begin{equation}
\Delta_{h,V}=\frac{\Delta_{h,F}(B)^{3}}{18\Delta_{lh}}\:,\label{eq:Voigt}
\end{equation}
where $\Delta_{h,F}(B)$ is the giant Zeeman splitting of holes in the Faraday geometry
\begin{equation}
\Delta_{h,F}=6\mathcal B= xN_{0}\beta \langle S^{Mn} (B) \rangle.
\end{equation}
According to Eq.~\eqref{eq:Voigt}, the Zeeman splitting of heavy holes is suppressed in the Voigt geometry as compared to the Faraday case: it is cubic in magnetic field at low fields and requires an admixture of light hole states. At high magnetic fields, $\mathcal B\gg \gamma_{1,2}k_{z}^{2}$, the angular momentum of the valence band states is oriented along the magnetic field, $\bm J_{h}\parallel x$. The energy spectrum is given by $E_{h}=\pm 3\mathcal B,E_{h}=\pm \mathcal B$, see also Fig.~\ref{fig:SMagn}(a). From now on we will label the valence band states by the corresponding projections of the hole momentum $J_{h,x}$ in the limit $\mathcal B \to +\infty$. Taking into account that $\beta<0$, the upper valence band state is $\psi_{h,-3/2},$ and the first excited state is $\psi_{h,-1/2}$.

Now we turn to the analysis of $\sigma$-type optical transitions between the Zeeman-split ground doublets of conduction and valence band states ($\pi$-type transitions remain $x$-polarized and are not relevant for the transverse routing effect). The optically allowed $\psi_{e\uparrow}\to \psi_{h,-1/2}$ and $\psi_{e\downarrow}\to \psi_{h,-3/2}$ transitions are
shown by the blue and red arrows in Fig.~\ref{fig:SMagn}(a). The magnetic field induces circular polarization of these transitions. Namely, at zero magnetic field, the heavy hole states include Bloch functions of $X$ and $Y$ symmetry only, and hence correspond to in-plane oscillations of the exciton dipole moment~\cite{Ivchenko}, so that the circular effect is absent. Nonzero magnetic field leads to admixture of the light hole states with nonzero contribution of the $Z$-type Bloch function to the heavy hole states. As a result, the exciton dipole moment acquires nonzero circular polarization. In order to quantify this effect we introduce the circular polarization degree of the exciton states at finite magnetic field,
\begin{equation}
\begin{gathered}
P_{c,\uparrow (\downarrow)}=\frac{I_{+,\uparrow (\downarrow)}-I_{-,\uparrow (\downarrow)}}{I_{+,\uparrow (\downarrow)}+I_{-,\uparrow (\downarrow)}},\\
I_{\pm,\uparrow}=| \langle \psi_{h,-1/2} |(p_{y}\pm {\rm i} p_{z})|\psi_{e\uparrow}\rangle|^{2},\quad
I_{\pm,\downarrow}=| \langle \psi_{h,-3/2}|(p_{y}\pm {\rm i} p_{z})|\psi_{e\downarrow}\rangle|^{2}
\:,
\end{gathered}
\end{equation}
where $\bm p$ is the momentum operator. We do not take into account the deviations from cubic symmetry so that the only nonzero momentum matrix element is $p_{cv}=\langle s|p_{y}|Y\rangle=\langle s|p_{z}|Z\rangle$. In this approximation, diagonalizing the valence band Hamiltonian \eqref{eq:HB} and substituting the explicit expressions for the $J_{h,x}=3/2,-1/2$ states from Ref.~\cite{Ivchenko}, we obtain the following results for the circular polarization degree:
\begin{equation}
\begin{gathered}
P_{c,\uparrow}\approx \frac{2(9-8Z^2+4ZZ_{-}+12Z-3Z_{-})}{16Z^2-8ZZ_{-}-24Z+6Z_{-}+63},\\
P_{c,\downarrow}\approx \frac{2(-9+8Z^2+4ZZ_{+}+12Z+3Z_{+})}{16Z^2+8ZZ_{+}+24Z+6Z_{+}+63},\\
Z=\frac{\Delta_{h,F}}{\Delta_{lh}},\quad Z_{\pm}=\sqrt{4Z^{2}\pm 6Z+9}\:.\label{eq:Pc}
\end{gathered}
\end{equation}
At large magnetic fields when the hole angular momentum is along the field direction, both polarizations
$P_{c,\uparrow}$ and $P_{c,\downarrow}$ saturate at unity.
Their dependence on magnetic field is shown in Fig.~\ref{fig:SMagn}(b) by the blue and red solid curves.
It is crucial that both transitions have the same sign of the circular polarization degree, because the angular momentum projection differences $S_{e,x}-J_{h,x}=1/2-(-1/2)=(-1/2)-(-3/2)=1$ have the same sign. Hence, the contributions of two lowest bright exciton states to the plasmon routing effects add up and do not cancel each other.
 At low magnetic fields we obtain
\begin{equation}
P_{c,\uparrow}\approx P_{c,\downarrow}\approx \frac{2}{3}\frac{\Delta_{h,F}}{\Delta_{lh}}\propto B\:,\label{eq:Pca}
\end{equation}
i.e. the circular polarization degree of the heavy-hole excitons in the Voigt geometry is linear in magnetic field. This result is somewhat counterintuitive since the Zeeman splitting in the Voigt geometry Eq.~ \eqref{eq:Voigt} is cubic in magnetic field at low fields. The approximate expression Eq.~\eqref{eq:Pca} is shown by the black dotted curve in Fig.~\ref{fig:SMagn}(b) and agrees well with the full results Eq.~\eqref{eq:Pc}.
We note, that the results Eq.~\eqref{eq:Pc}, \eqref{eq:Pca} were obtained in the approximation of infinite QW barriers. The finite height of the barriers can be taken into account by multiplying Eq.~\eqref{eq:Pca} by the squared overlap integral of the light and heavy hole envelope functions, $|\langle\psi_{hh}|\psi_{lh}\rangle|^{2}$, which decreases the polarization degree.\\


\noindent{\bf 3. Sample preparation}

The semiconductor QW structures were grown by molecular-beam epitaxy, which allows achieving high quality structures with well-defined layer arrangement and sub-nanometer thickness precision. The 10~nm thick diluted-magnetic-semiconductor Cd$_{0.95}$Mn$_{0.05}$Te QW layer ($E_g=1.656$~eV at $2\,$K) was grown after deposition of a 3~$\mu$m thick Cd$_{0.73}$Mg$_{0.27}$Te buffer layer ($E_g \approx 2.10$~eV at $2\,$K) on a (100) GaAs substrate. The QW was covered with a thin Cd$_{0.73}$Mg$_{0.27}$Te cap layer, which serves as an electronic barrier for the excitons in the QW and simultaneously as a spacer between the QW layer and the structure's surface. Two structures with different thicknesses of the cap layers, 250 nm and 32 nm, referred to as sample 1 and sample 2, respectively, were used in the studies.

On top of the semiconductor structure $200\times200~\mu{\rm m}^2$ sized gold gratings were patterned, using electron beam lithography and "lift-off" processing. The used electron beam lithography instrument was the Raith VOYAGER system. In the "lift-off" processing no metal was used to promote the adhesion of the gold layer. Instead the sample covered with a PMMA resist mask was subjected to a short plasma treatment for descumming directly in the metal evaporation chamber, right before the evaporation of gold. The grating period, slit width, and gold thickness were $250\,$nm, $55\,$nm, and $45\,$nm respectively. The heterostructure was also studied without gold at the top, which we refer to as the bare QW.\\

\noindent{\bf 4. Experimental setup}

For optical studies we use a Fourier imaging setup in combination with a spectrometer, which allows us to record the angular and spectral dependence of the emitted intensity in a single acquisition step. In PL imaging measurements we use laser excitation with photon energy of 2.25~eV in order to populate the QW with excitons. On the way to the sample the laser light passes a $50:50$ non-polarizing beam splitter cube and is subsequently linearly polarized by a Glan-Thompson prism. Subsequently it is focused onto the sample using a $20\times$ microscope objective with a numerical aperture $0.4$ into a spot of about $20\,\mu$m diameter. The excitation density (below $50$~W/cm$^2$) is kept low enough to avoid heating of the Mn system. The sample was mounted on the cold finger of a liquid Helium flow cryostat and was kept at a temperature of about $10\,$K. Transverse magnetic fields up to $520\,$mT were applied in the QW sample plane using an electromagnet ($B \parallel x$).

The photoluminescence from the QW was collected using the same microscope objective in backscattering geometry. At a distance of twice the focal length of the objective the Fourier plane is located. In this plane the vertical position inside the emitted light cone corresponds to an emission angle of $-23^\circ$ to $23^\circ$. After passing the Glan-Thompson polarizer, the emitted light is deflected inside the non-polarizing beam splitter cube. A pair of achromatic lenses subsequently maps the Fourier plane onto the slit of the spectrometer. The stray light from the excitation laser was suppressed by a $600\,$nm long pass filter in front of the spectrometer. While the angular information was stored in the vertical direction, the emitted light was spectrally dispersed ($6.4\,$nm/mm) in horizontal direction using a single imaging monochromator and detected by a charge-coupled device camera with $256\times1024\,$pixels respectively, resulting in an angular resolution of $\le0.5^\circ$ and a spectral resolution of $0.6\,$nm.

\begin{figure}[t]
\centering
\includegraphics[width=\textwidth]{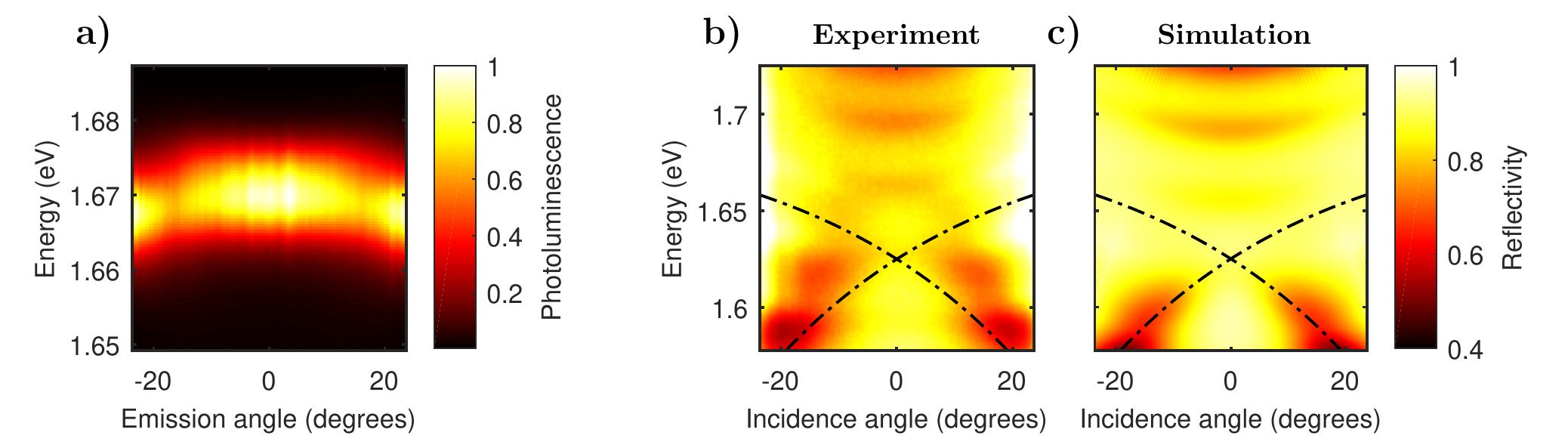}
\caption{\textbf{Optical properties of the studied hybrid structures: Photoluminescence and reflectivity spectra.} (a) Angle- and spectrally-resolved PL from the QW in sample 1 under off-resonant excitation with photon energy 2.25~eV. (b,c) Contour plots of angle-resolved experimental (b) and calculated (c) reflectivity spectra of Sample 1. The interference fringes originate from the finite buffer thickness. The dispersion of the SPPs is shown by the dashed dotted lines following the maximum of absorption in the calculated spectra. PL and reflectivity images are measured at $T=10$~K.}
\label{Fig:Experiment}
\end{figure}

An examplary contour plot of the PL intensity $I(\hbar\omega,\theta)$ taken on sample 1 is presented in Fig.~\ref{Fig:Experiment}(a). The emission band of the QW excitons is centered at the photon energy $\hbar\omega \approx 1.67$~eV and depends weakly on the emission angle $\theta$. The spectral width  at half maximum of the PL band of about 10~meV allows us to measure the emission pattern in a spectral range from 1.65 to 1.69~eV.

Reflectivity spectra provide important information about the positions of the surface plasmon-polariton (SPP) resonances and their dispersion. Fig.~\ref{Fig:Experiment}(b) shows the angle- and spectrally-resolved reflectivity measured on sample 1 at $T=10$~K for $p$-polarized light. The broad dips in the reflectivity spectra are attributed to SPP resonances at the semiconductor-gold interface. The spectral positions of these features strongly depend on the incidence angle and are absent in $s$-polarized spectra and if taken from bare QW structures. The dash-dotted lines show the dispersion of the corresponding SPPs. The fast oscillations with a period of about 25~meV arise from interference fringes, which originate from the finite buffer thickness. The experimental data are in good agreement with the calculated reflectivity spectra (Fig.~\ref{Fig:Experiment}(c)) which demonstrate the same features. The rigorous electromagnetic simulations were performed using the structural parameters given above.\\

\noindent{\bf 5. Electromagnetic simulations}

The PL and reflection spectra were calculated using the scattering matrix method from Ref.~\cite{Whittaker}. In each layer of the structure the electromagnetic field is decomposed into plane waves, that are coupled due to Bragg diffraction. The Li factorization technique \cite{Li} was used to improve the convergence of the Fourier series. It was sufficient to take into account $2h_{\rm max}+1=31$ reciprocal lattice vectors to achieve convergence. In the PL calculations we modeled excitons as uncorrelated point dipoles with fixed polarization, randomly distributed in the $xy$ plane. The values of the frequency-dependent refractive indices for gold, CdMgTe and GaAs were taken from Ref.~\cite{JohnsonChristy}, Ref.~\cite{AndreDang} and Ref.~\cite{Aspnes}, respectively. The background refractive index contrast between the QW and the CdMgTe spacer layer was neglected.



\begin{thebibliography}{1}

\bibitem{Benson-2011} O.~Benson, {\it Assembly of hybrid photonic architectures from nanophotonic constituents}, Nature (London) {\bf 480}, 193 (2011).

\bibitem{Lodahl-et-al-2017} P.~Lodahl, S. Mahmoodian, S. Stobbe, A. Rauschenbeutel, P. Schneeweiss, J. Volz, H. Pichler and P. Zoller, {\it Chiral quantum optics}, Nature (London) {\bf 541}, 473 (2017).

\bibitem{Temnov-2010} V.V.~Temnov, G.~Armelles, U.~Woggon, D.~Guzatov, A.~Cebollada, A. Garcia-Martin, J.-M. Garcia-Martin, T. Thomay, A.~Leitenstorfer, and R. Bratschitsch, {\it Active magneto-plasmonics in hybrid metal�ferromagnet structures}, Nat. Photonics {\bf 4}, 107 (2010).

\bibitem{Akimov-2012} I.A.~Akimov, V.I.~Belotelov, A.V.~Scherbakov, M.~Pohl, A.N.~Kalish, A.S.~Salasyuk, M.~Bombeck, C.~Brueggemann, A.V.~Akimov, R.I.~Dzhioev, V.L.~Korenev, Yu.G.~Kusrayev, V.F.~Sapega, V.A.~Kotov, D.R.~Yakovlev, A.K.~Zvezdin, and M. Bayer, {\it Hybrid structures of magnetic semiconductors and plasmonic crystals: A novel concept for magneto-optical devices}, J. Opt. Soc. Am. B {\bf 29}, A103 (2012).

\bibitem{Armelles-2012} G. Armelles, A. Cebollada, A. Garc\'{\i}a-Mart\'{\i}n, and M. U. Gonz\'{a}lez, {\it Magnetoplasmonics: Combining magnetic and plasmonic functionalities}, Adv. Opt. Mater. {\bf 1}, 10 (2013).

\bibitem{Bossini-2016} D.~Bossini, V.I.~Belotelov, A.K.~Zvezdin, A.N.~Kalish, and A.V.~Kimel, {\it Magnetoplasmonics and Femtosecond Optomagnetism at the Nanoscale}, ACS Photonics {\bf 3}, 1385 (2016).

\bibitem{Zvezdin-TMOKE} Zvezdin, A. \& Kotov, V. Modern Magnetooptics and Magnetooptical Materials (IOP, 1997).

\bibitem{MO-book} Magnetophotonics: From theory to applications, Eds: M.~Levy, A.V.~Baryshev, M.~Inoue (Springer Verlag, Berlin, Heidelberg 2013).

\bibitem{Belotelov-2009} V.I.~Belotelov, I.A.~Akimov, M.~Pohl, V.A.~Kotov, S.~Kasture, A.S.~Vengurlekar, A. Venu Gopal, D.R.~Yakovlev, A.K. Zvezdin, and M. Bayer, {\it Enhanced magneto-optical effects in magnetoplasmonic crystals}, Nat. Nanotech. {\bf 6}, 370 (2011).

\bibitem{Chin-2013} J.Y.~Chin, T.~Steinle, T. Wehlus, D.~Dregely, T. Weiss, V.I. Belotelov, B. Stritzker, and H. Giessen, {\it Nonreciprocal plasmonics enables giant enhancement of thin-film Faraday rotation}, Nat. Commun. {\bf 4}, 1599 (2013)

\bibitem{Kreilkamp-2013} L.E.~Kreilkamp, V.I.~Belotelov, J.Y.~Chin, S.~Neutzner, D.~Dregely, T.~Wehlus, I.A.~Akimov, M.~Bayer, B.~Stritzker, and H.~Giessen, {\it Waveguide-plasmon polaritons enhance transverse magneto-optical Kerr effect}, Phys. Rev. X {\bf 3}, 041019 (2013).

\bibitem{Floess-2017} D.~Floess, M.~Hentschel, T.~Weiss, H.-U. Habermeier, J.~Jiao, S.G.~Tikhodeev, and H.~Giessen, {\it Plasmonic analog of electromagnetically induced absorption leads to giant thin film Faraday rotation of $14^{\circ}$}, Phys. Rev. X {\bf 7}, 021048 (2017).

\bibitem{MChD97} G.L.J.A.~Rikken and E.~Raupach, {\it Observation of magneto-chiral dichroism}, Nature {\bf 390}, 493 (1997).

\bibitem{Bliokh-2015} A. Y. Bekshaev, K. Y. Bliokh and F. Nori, {\it Transverse Spin and Momentum in Two-Wave Interference}, Phys. Rev. X {\bf 5}, 011039 (2015).


\bibitem{Glazov-2007} C. Leyder, M. Romanelli, J. P. Karr, E. Giacobino, T. C. H. Liew, M. M. Glazov, A. V. Kavokin, G. Malpuech and A. Bramati, {\it Observation of the optical spin Hall effect}, Nature Physics {\bf 3}, 628 (2007).

\bibitem{Bliokh-et-al-2015} K. Y. Bliokh, F.J. Rodr\'iguez-Fortu\~no, F. Nori and A. V. Zayats, {\it Spin-orbit interactions of light}, Nat. Photon. {\bf 9}, 796 (2015).

\bibitem{Zayatz-2015} F.J.~Rodr\'iguez-Fortu\~no, N.~Engheta, A. Mart\'inez, and A.V.~Zayats, {\it Lateral forces on circularly polarizable particles near a surface}, Nat. Commun. {\bf 6}, 8799 (2015).

\bibitem{Poddubny-2014} P.V.~Kapitanova, P.~Ginzburg, F.J.~Rodr\'iguez-Fortu\~no, D.S.~Filonov, P.M. Voroshilov, P.A.~Belov, A.N.~Poddubny, Yu.S. Kivshar, G.A.~Wurtz, and A.V.~Zayats, {\it Photonic spin Hall effect in hyperbolic metamaterials for polarization-controlled routing of subwavelength modes}, Nat. Commun. {\bf 5}, 3226 (2014).

\bibitem{Leuchs-rev-2015} A.~Aiello, P.~Banzer, M.~Neugebauer and G.~Leuchs, {\it From transverse angular momentum to photonic wheels}, Nat. Photon. {\bf 9}, 789 (2015).

\bibitem{Rauschenbeutel-2013} C.~Junge, D.~O'Shea, J.~Volz, and A.~Rauschenbeutel, {\it Strong coupling between single atoms and nontransversal photons}, Phys. Rev. Lett. {\bf 110}, 213604 (2013).

\bibitem{Dayan-2014} I.~Shomroni, S.~Rosenblum, Y.~Lovsky, O.~Bechler, G.~Guendelman, and B.~Dayan, {\it  All-optical routing of single photons by a one-atom switch controlled by a single photon}, Science {\bf 345}, 903 (2014).

\bibitem{Oulton-2013} I.J.~Luxmoore, N.A.~Wasley, A.J.~Ramsay, A.C.T.~Thijssen, R.~Oulton, M.~Hugues, S.~Kasture, V. G. Achanta, A.M.~Fox, and M.S.~Skolnick, {\it Interfacing spins in an InGaAs quantum dot to a semiconductor-waveguide circuit using emitted photons }, Phys. Rev. Lett. {\bf 110}, 037402 (2013).

\bibitem{Lodahl-2015} I. S\"ollner, S. Mahmoodian, S. Lindskov Hansen, L. Midolo, A. Javadi, G. Kir\v{s}ansk\.{e}, T. Pregnolato, H. El-Ella, E.H. Lee, J.D. Song, S. Stobbe and P. Lodahl, {\it Deterministic photon emitter coupling in chiral photonic circuits}, Nat. Nanotechnology {\bf 10}, 775 (2015).

\bibitem{Makhonin-2016} R.J.~Coles, D.M.~Price, J.E.~Dixon, B.~Royall, E.~Clarke, P.~Kok, M.S.~Skolnick, A.M.~Fox, and M.N.~Makhonin, {\it Chirality of nanophotonic waveguide with embedded quantum emitter for unidirectional spin transfer}, Nat. Commun. {\bf 7}, 11183 (2016).

\bibitem{Capasso-2013} J.~Lin, J.P.~Balthasar Mueller, Q.~Wang, G.~Yuan, N.~Antoniou, X.-C.~Yuan, F.~Capasso, {\it Polarization-controlled tunable directional coupling of surface plasmon polaritons}, Science {\bf 340}, 331 (2013).

\bibitem{Zayats-2013} F.J. Rodr\'{\i}guez-Fortu\~{n}o, G.~Marino, P.~Ginzburg, D.~O\'Connor, A.~Mart\'{\i}nez, G.A.~Wurtz, and A.V.~Zayats, {\it Near-field interference for the unidirectional excitation of electromagnetic guided modes}, Science {\bf 340}, 328 (2013).

\bibitem{Leuchs-2013} T.~Bauer, S.~Orlov, U.~Peschel, P.~Banzer, and G.~Leuchs, {\it Nanointerferometric amplitude and phase reconstruction of tightly focused vector beams}, Nat. Photon. {\bf 8}, 23 (2013).

\bibitem{Zayats-2014} D.~O'Connor, P. Ginzburg, F.J. Rodr\'iguez-Fortu\~no, G.A. Wurtz and A.V. Zayats, {\it Spin-orbit coupling in surface plasmon scattering by nanostructures}, Nat. Commun. {\bf 5}, 5327 (2014).

\bibitem{Rauschenbeutel-2014} J.~Petersen, J.~Volz, and A.~Rauschenbeutel, {\it Chiral nanophotonic waveguide interface based on spin-orbit interaction of light}, Science {\bf 346}, 67 (2014).

\bibitem{Martinez-2014} F.J.~Rodr\'{\i}guez-Fortu\~{n}o, I.~Barber-Sanz, D.~Puerto, A.~Griol, and A.~Mart\'{\i}nez, {\it Resolving light handedness with an on-chip silicon microdisk }, ACS Photonics {\bf 1}, 762 (2014).

\bibitem{Grosjean-2015} Y. Lefier and T. Grosjean, {\it Unidirectional sub-diffraction waveguiding based on optical spin�orbit coupling in subwavelength plasmonic waveguides}, Optics Lett. {\bf 40}, 2890 (2015).

\bibitem{Iorsh-2017} I.S. Sinev, A.A. Bogdanov, F.E. Komissarenko, K.S. Frizyuk, M.I. Petrov, I.S. Mukhin, S.V. Makarov, A.K. Samusev, A.V. Lavrinenko, I.V. Iorsh, {\it Dielectric nanoantenna as an efficient and ultracompact demultiplexer for surface waves}, arXiv:1705.07689 (2017).

\bibitem{supplement} Supplementary Material: Section 1 describes the Zeeman splitting of exciton states in Faraday geometry; Section 2 describes the selection rules for optical transitions and the polarization degree of exciton dipoles in transverse magnetic field (Voigt geometry); Section 3 - details on sample preparation; Section 4 - experimental setup and optical properties of the investigated samples; Section 5 - Electromagnetic simulations.

\bibitem{Furdyna} J.K.~Furdyna, {\it Diluted magnetic semiconductors}, J. Appl. Phys. {\bf 64}, R29 (1988).

\bibitem{Kossut-book}
J.A.~Gaj and  J.~Kossut, {\it Introduction to the Physics of Diluted Magnetic Semiconductors}, (Springer, Berlin, Heidelberg 2010).

\bibitem{DMS-1994} B.~Kuhn-Heinrich, W.~Ossau, E.~Bangert, A.~Waag, and G.~Landwehr, {\it Zeeman pattern of semimagnetic (CdMn)Te/(CdMg)Te quantum wells in inplane magnetic fields}, Solid State Comm. {\bf 91}, 413 (1994).












\end{thebibliography}

\begin{thebibliography}{1}

\bibitem{Furdyna} J.K.~Furdyna, {\it Diluted magnetic semiconductors}, J. Appl. Phys. {\bf 64}, R29 (1988).

\bibitem{Kossut-book}
J.A.~Gaj, J.~Kossut, {\it Introduction to the Physics of Diluted Magnetic Semiconductors}, (Springer, Berlin, Heidelberg, 2010).

\bibitem{DMS-1994} B.~Kuhn-Heinrich, W.~Ossau, E.~Bangert, A.~Waag, and G.~Landwehr, {\it Zeeman pattern of semimagnetic (CdMn)Te/(CdMg)Te quantum wells in inplane magnetic fields}, Solid State Comm. {\bf 91}, 413 (1994).

\bibitem{Winkler}R. Winkler, {\it Spin-orbit Coupling Effects in Two-Dimensional Electron and Hole Systems} (Springer, 2003).

\bibitem{LeSiDang} L.S. Dang, G. Neu, and R. Romestain, {\it Optical detection of cyclotron resonance of electron and holes in CdTe}, Solid State Commun. {\bf 44}, 1187 (1982).

\bibitem{Strain} H. Mariette, F. Dalbo, N. Magnea, G. Lentz, and H. Tuffigo, {\it Optical investigation of confinement and strain effects in CdTe/(Cd,Zn)Te single quantum wells}, Phys. Rev. B {\bf 38}, 12443 (1988).

\bibitem{Yakovlev-92} E.L. Ivchenko, A.V. Kavokin, V.P. Kochereshko, G.R. Posina, I.N. Uraltsev, D.R. Yakovlev, R.N. Bicknell-Tassius, A. Waag, and G. Landwehr, {\it Exciton oscillator strength in magnetic-field-induced spin superlattices CdTe/(Cd,Mn)Te},  Phys. Rev. B {\bf 46}, 7713 (1992).

\bibitem{Ivchenko} E.L. Ivchenko, {\it Optical Spectroscopy of Semiconductor Nanostructures} (Alpha Science International, 2005).

\bibitem{Whittaker} D. M. Whittaker and I. S. Culshaw, {\it Scattering-matrix treatment of patterned multilayer photonic structures}, Phys. Rev. B {\bf 60}, 2610 (1999).

\bibitem{Li} L. Li, {\it Use of Fourier series in the analysis of discontinuous periodic structures}, J. Opt. Soc. Am. A {\bf 13}, 1870 (1996).

\bibitem{JohnsonChristy} P. B. Johnson and R. W. Christy, {\it Optical Constants of the Noble Metals}, Phys. Rev. B {\bf 6}, 4370 (1972).

\bibitem{AndreDang} R. Andre and L. S. Dang, {\it Low-temperature refractive indices of Cd$_{1-x}$Mn$_{x}$Te and Cd$_{1-x}$Mg$_{x}$Te}, J. Appl. Phys. {\bf 82}, 5086 (1997).

\bibitem{Aspnes} D. E. Aspnes, S. M. Kelso, R. A. Logan and R. Bhat, {\it Optical properties of Al$_{x}$Ga$_{1-x}$As}, J. App. Phys. {\bf 60}, 754-767 (1986).





\end{thebibliography}
\end{document}